\documentclass[12pt,preprint]{aastex}
\usepackage{natbib,graphicx}
\bibpunct[, ]{(}{)}{;}{a}{}{,}
\def \aj {AJ}
\def \mnras {MNRAS}
\def \pasp {PASP}
\def \apj {ApJ}
\def \apjs {ApJS}
\def \apjl {ApJL}
\def \aap {A\&A}
\def \nat {Nature}

\def \iaucirc {IAUC}

\newcommand{\msol} {M$_{\odot}$}

\def\lesssim{\mathrel{\hbox{\rlap{\hbox{\lower4pt\hbox{$\sim$}}}\hbox{$<$}}}}
\def\gtrsim{\mathrel{\hbox{\rlap{\hbox{\lower4pt\hbox{$\sim$}}}\hbox{$>$}}}}
\begin{document}
%%%%%%%%%%%%%%%%%%%%%%%%%%%%%%%
%Title Page                   %
%%%%%%%%%%%%%%%%%%%%%%%%%%%%%%%
\title{Luminosity and mass limits for the progenitor of the type Ic supernova 2004gt in NGC 4038}
\shorttitle{The Progenitor of SN 2004gt}
\shortauthors{Maund, Smartt and Schweizer}

\author{Justyn R.~Maund\footnote{jrm@ast.cam.ac.uk, Institute of Astronomy, University of Cambridge, Madingley Road, Cambridge CB3 0HA, U.K.} ~and Stephen J. ~Smartt\footnote{Department of Physics and Astronomy, Queen's University Belfast, Belfast, BT7 1NN, U.K.} ~and Francois~Schweizer\footnote{Carnegie Observatories, 813 Santa Barbara Street, Pasadena, California 91101 USA}}
%%%%%%%%%%%%%%%%%%%%%%%%%%%%%%%%
\begin{abstract}
We report our attempts to locate the progenitor of the type Ic SN 2004gt in NGC 4038.  We use high resolution HST ACS images of SN 2004gt and have compared them with deep pre-explosion HST WFPC2 F336W, F439W, F555W and F814W images.  We identify the SN location on the pre-explosion frames with an accuracy of 5mas.  We show that the progenitor is below the detection thresholds of all the pre-explosion images.  These detection limits are used to place luminosity and mass limits on the progenitor.  The progenitor of SN 2004gt seems to be restricted to a low-luminosity high-temperature star, either a single WC star with initial mass $>40M_{\odot}$ or a low mass star in a binary.  The pre-explosion data cannot distinguish between the two scenarios.
\end{abstract}
\maketitle
%%%%%%%%%%%%%%%%%%%%%
%Introduction
%%%%%%%%%%%%%%%%%%%%
\section{INTRODUCTION}
\label{intro}
Stellar-evolution models predict that all stars with $M_{ZAMS}\gtrsim8M_{\odot}$ should end their lives in a core-collapse induced supernova (CCSN) explosion (e.g. \citet{heg03,eld04}).  The identification of the progenitor of a CCSN just prior to explosion provides a critical test for models of stellar evolution. The task of finding the progenitors is complicated by the unpredictable nature of SNe. The two nearby supernovae SN~1987A \citep{walb87a} and SN~1993J \citep{alder93j} had progenitors identified shortly after discovery.  Modern telescope archives have allowed the search for progenitors to be increased to large distances \citep{2005astro.ph..1323M,vandykprog}. In order to confidently identify the correct star as a SN progenitor on images of the highest resolution, precise astrometric techniques are required. Recently, \citet{2005astro.ph..1323M} and \citet{smartt03gd} have used high resolution post-explosion images with the Hubble Space Telescope (HST) and the technique of differential astrometry to accurately identify progenitors on pre-explosion HST images. The brightness and colour information from pre-explosion images may then be used to place the progenitor on the Hertzsprung-Russell (HR) diagram (e.g SN 2003gd, \citealt{smartt03gd}).  There are no detections yet of progenitors of the hydrogen deficient SNe Ib or Ic, although upper limits to their absolute magnitudes have been determined (see \S\ref{disc}). It is now thought that some SNe Ic produce gamma-ray bursts (GRBs; \citealt{2003ApJ...599..394M,hjo03dh}), and that the progenitors may be rapidly rotating massive Wolf-Rayet (WR) stars.\\
This letter presents the study of HST Wide Field Planetary Camera 2 (WFPC2) images of the site of SN 2004gt in the interacting galaxy NGC 4038, prior to explosion.  SN 2004gt was discovered by L.A.G. Monard \citep{2004IAUC.8454....1M} on 2004 Dec 12.076 with an unfiltered magnitude of 14.6, corresponding to an absolute magnitude of $\approx -17$ at
the distance of NGC 4038 ($(m-M)_{0}=31.4$; \citealt{1999AJ....118.1551W}).  The position of SN 2004gt was given as $\alpha_{2000}=12^{h}01^{m}50^{s}.37$, $\delta_{2000}=-18\degr52\arcmin12.\arcsec7$.  \citet{2004IAUC.8456....3G} classified SN 2004gt of Type Ic.  The WFPC2 data are the deepest, and widest wavelength coverage, pre-explosion images of any nearby SNe Ibc to date, and provide a unique opportunity to study the progenitor site of a possible GRB-related supernova event.  \citet{gal} use post-explosion ground-based imaging with Adaptive Optics, and similar pre-explosion HST WFPC2 data, and identify the same location for the progenitor of SN 2004gt and draw similar conclusions about its nature to those presented here.
\section{OBSERVATIONS AND DATA ANALYSIS} 
The pre-explosion HST WFPC2 observations of the site of SN 2004gt were obtained on 1996 Jan 20 with F336W (4500s), F439W (4000s), F555W (4400s) and F814W (2000s) filters.  The pre-explosion WFPC2 observations (Programme G0-5962; PI: B. Whitmore), in each filter, are drizzled combinations of four separate exposures.  These four exposures were acquired in pairs, for the rejection of cosmic rays, with an offset of $0.25\arcsec$ between pairs of exposures \citep{1999AJ....118.1551W}.  The progenitor of SN 2004gt fell on the Planetary Camera (PC) chip for which the pixel size of the drizzled image was $0.025\arcsec$ (the same as the HST Advanced Camera for Surveys (ACS) High Resolution Channel (HRC)).   Aperture photometry was performed on the final images using the {\sc Iraf} implementation of {\sc DAOphot} and the prescription of \citet{1999AJ....118.1551W}.  Empirical aperture corrections were calculated for each frame and  small charge transfer inefficiency corrections ($\mathrm{\sim0.01mag}$) were applied to the photometry using the prescription of \citet{dolp00cte}.  The photometry was converted to the standard Johnson-Cousins system using the transformations of \citet{holsphot95} with the updates of \citet{dolphhstphot}.\\  
{\sc DAOphot} provides four centering algorithms to calculate the locations of stars: centroid, ofilter, Gaussian and PSF fitting.  The scatter in the measured locations of stars was used to describe the uncertainty due to the under-sampled PSF of WFPC2.  Post-explosion ACS/HRC images (Programme GO-10187; PI: S. Smartt) were acquired on 2005 May 16 with the F435W (1672s), F555W (1530s) and F814w (1860s) filters.  These images were analysed using the {\sc PyRaf} implementation of {\sc DAOphot}, using Tiny Tim PSFs \citep{tinytim04} for PSF photometry.  Empirical aperture corrections to $0.5\arcsec$ were calculated for each frame (with the correction for $0.5\arcsec$ to $\infty$ taken from \citet{acscoltran}), and the photometry was corrected for charge transfer inefficiency according to the relationships of \citet{riesscte}. The ACS photometry was converted to standard Johnson-Cousins magnitudes using the transformations of \citet{acscoltran}.  The photometry was found to agree with results from the {\sc DOLPhot} package, an update of {\sc HSTPhot} \citep{dolphhstphot} to $\lesssim0.1$ mag at $m\approx25$.  The locations of 16 stars common to both the pre-explosion WFPC2 imaging and the post-explosion ACS imaging were used to calculate the transformation between the two sets of F555W images, using the {\sc Iraf} task {\it geomap}.  The positions of stars on the ACS frame were transformed to the coordinates of the drizzled PC F555W image with an accuracy of $\mathrm{\pm5}$mas.  The reverse transformation has an uncertainty of $\mathrm{\pm4.6}$mas.  SN 2004gt is located $1.3\arcsec$, approximately South, from knot S\citep{1999AJ....118.1551W}.
\section{OBSERVATIONAL RESULTS}
\label{results}
\begin{figure}
\rotatebox{-90}{\includegraphics[width=5cm]{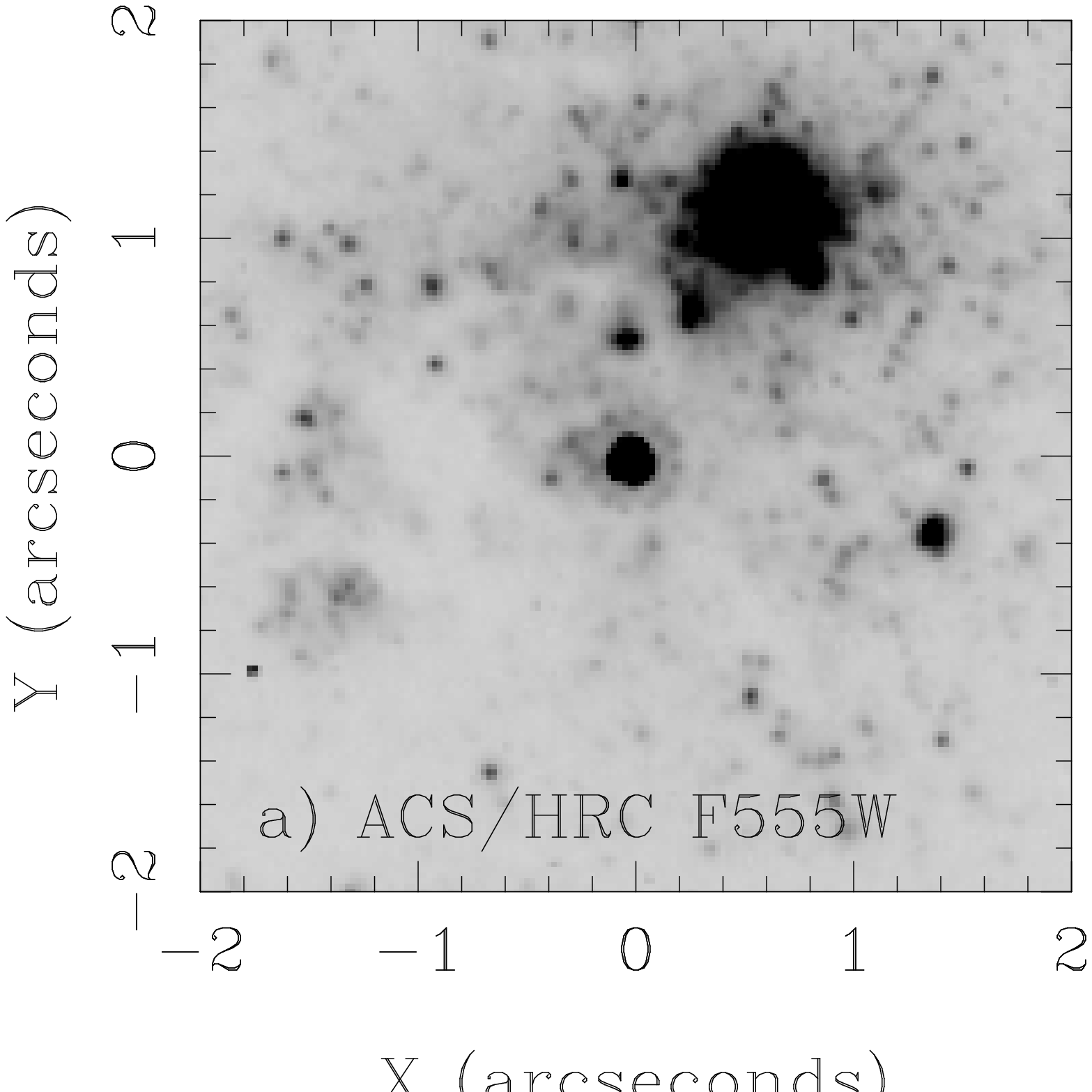}}
\rotatebox{-90}{\includegraphics[width=5cm]{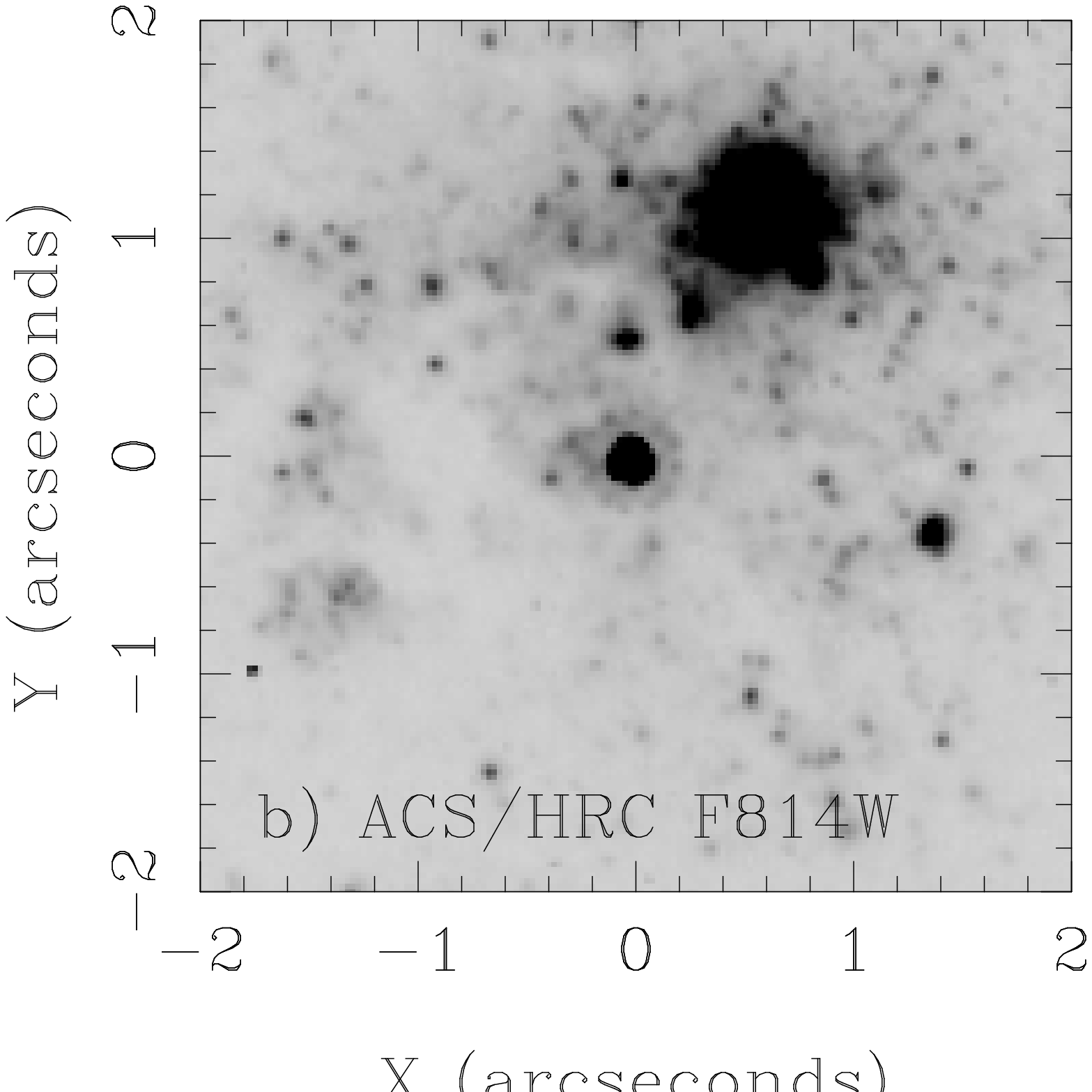}}
\rotatebox{-90}{\includegraphics[width=5cm]{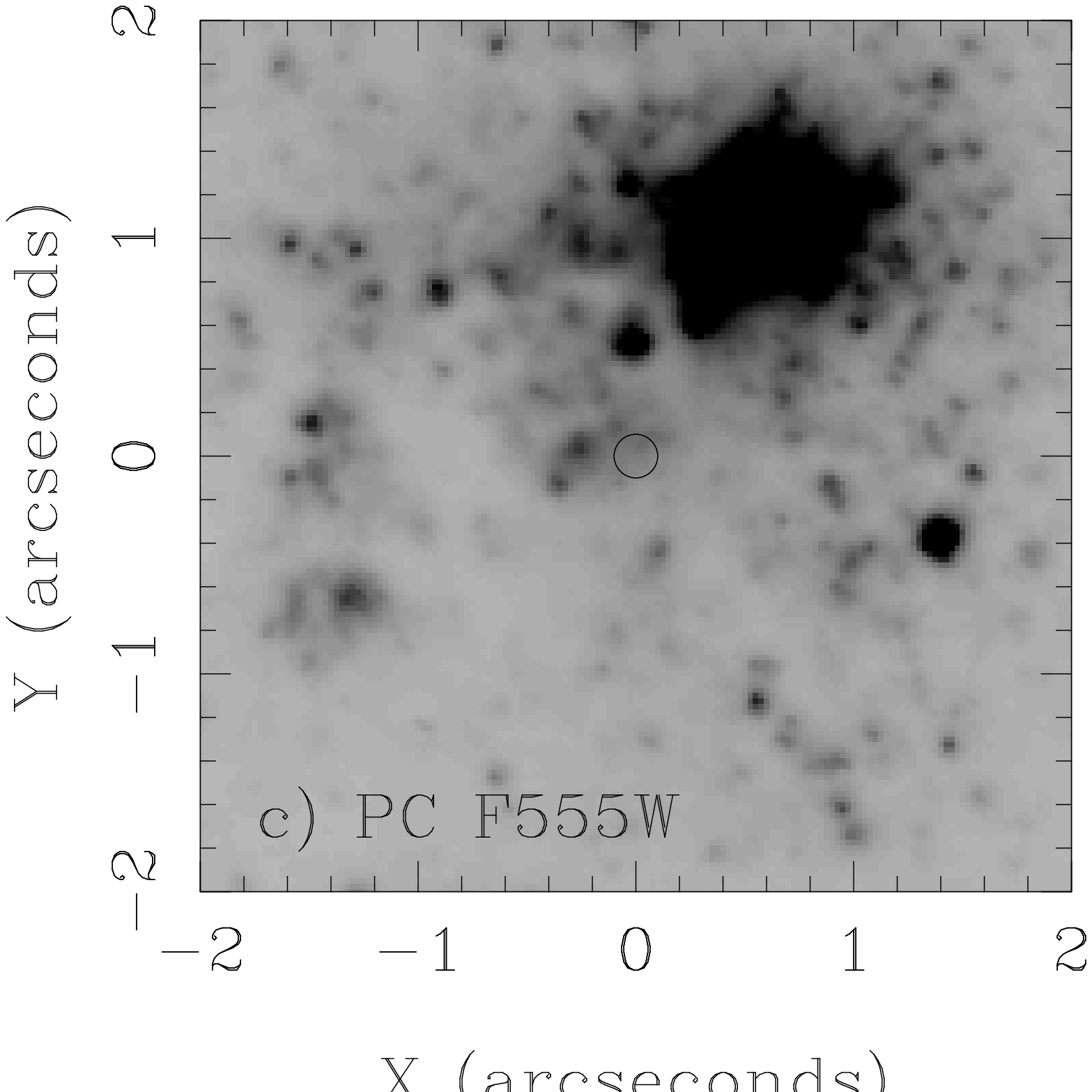}}
\caption{Pre- and Post-explosion images of the site of SN 2004gt in
NGC 4038.  The same orientation is used for all three panels, with North up and East to the left.  SN 2004gt is located at the centre of each of the panels.  a) Post-explosion ACS/HRC F555W image, with SN 2004gt clearly visible. b) Post-explosion ACS/HRC F814W image. c) Pre-explosion WFPC2 PC chip F555W image, transformed, shifted and rotated to the coordinates of the post-explosion ACS imaging.  No object is significantly detected at the position of SN 2004gt prior to explosion, with a positional uncertainty $0.004\arcsec$.  In these images knot S appears in the top-right.}
\label{04gtdiff} 
\end{figure}

SN 2004gt is readily identifiable on post-explosion ACS images.  We measure the apparent brightness and colours of SN 2004gt on 2005 May 16 as $\mathrm{m_{V}=18.55\pm0.12}$, $B-V=0.87\pm0.16$ and $V-I=1.32\pm0.15$.  The location of SN 2004gt and its position on the WFPC2 pre-explosion images are shown as Fig. \ref{04gtdiff}.  The limiting magnitude at the position of SN 2004gt on the pre-explosion frames was determined using the technique of \citet{2005astro.ph..1323M}.  Aperture photometry was done over a 9-point grid centered at the SN position, with offsets of 1 pixel which permitted background variations, over the aperture area, to be taken into account.  The detection thresholds determined in this manner were tested by adding a scaled PSF at the SN position on the pre-explosion magnitude.  This tested if the {\sc DAOphot} photometry would recover a star at the calculated detection threshold.  The detection thresholds were determined to be: $m_{F336W}\approx23.04$, $m_{F439W}\approx24.56$, $m_{F555W}\approx25.86$ and $m_{F814W}\approx24.43$.\\
The post-explosion ACS photometry of stars within $2\arcsec$ was used to estimate the reddening towards SN 2004gt using the technique of \citet{2005astro.ph..1323M}.  The photometry of 47 nearby stars was used to estimate the reddening as $E(B-V)=0.07\pm0.01$.  This low reddening is consistent with the foreground reddening as quoted by NED\footnote{http://nedwww.ipac.caltech.edu/}, after \citet{schleg98}, and the reddening determined by \citet{1999AJ....118.1551W} towards knot S.  The shape of the continuum of an early spectrum of SN 2004gt (Kinugasa, Private Communication) did not show evidence for significant reddening, being similar in shape to early time spectra of SN 2002ap ($E(B-V)=0.09$,\citealt{mazz02}).\\
The detection thresholds, calculated above, were converted to absolute magnitudes taking into account the distance and extinction towards NGC 4038.   We calculated the extinctions in WFPC2 bands, using the $A_{X}/E(B-V)$ relations of \citet{1999AJ....118.2331V}, to be: $A_{F336W}=0.39$, $A_{F439W}=0.30$, $A_{F555W}=0.20$ and $A_{F814W}=0.13$.  This implies the progenitor on the pre-explosion frames had magnitudes: $M_{F336W}\gtrsim-8.75$, $M_{F439W}\gtrsim-7.14$, $M_{F555W}\gtrsim-5.74$ and $M_{F814W}\gtrsim-7.10$.\\
An age for the environment around SN 2004gt was estimated by comparing the positions of nearby stars on a colour-magnitude diagram with theoretical isochrones (we use the stellar evolution tracks and isochrones of the Geneva stellar evolution group\footnote{http://webast.ast.obs-mip.fr/stellar/}).  The isochrones were shifted for the distance, extinction and reddening of NGC 4038.  Assuming the standard Galactic extinction-reddening relationship of $A_{V}\approx3.1E(B-V)$ and $E(V-I)=1.6E(B-V)$, the value of the reddening calculated above yielded $A_{V}=0.22\pm0.03$ and $E(V-I)=0.11\pm0.05$.  The mean age of the nearby stars was measured to be $\mathrm{log(age/years)=6.93\pm0.13}$, or $8.5\pm2.5$Myr.  This age is consistent with the age of $7\pm1$Myr estimated by \citet{1999AJ....118.1551W} for knot S.  The age of the nearby stars corresponds to the lifetime of stars with $M_{ZAMS}\approx20-40M_{\odot}$.
\section{DISCUSSION}
\label{disc}
\begin{figure}
\rotatebox{-90}{\includegraphics[width=5cm]{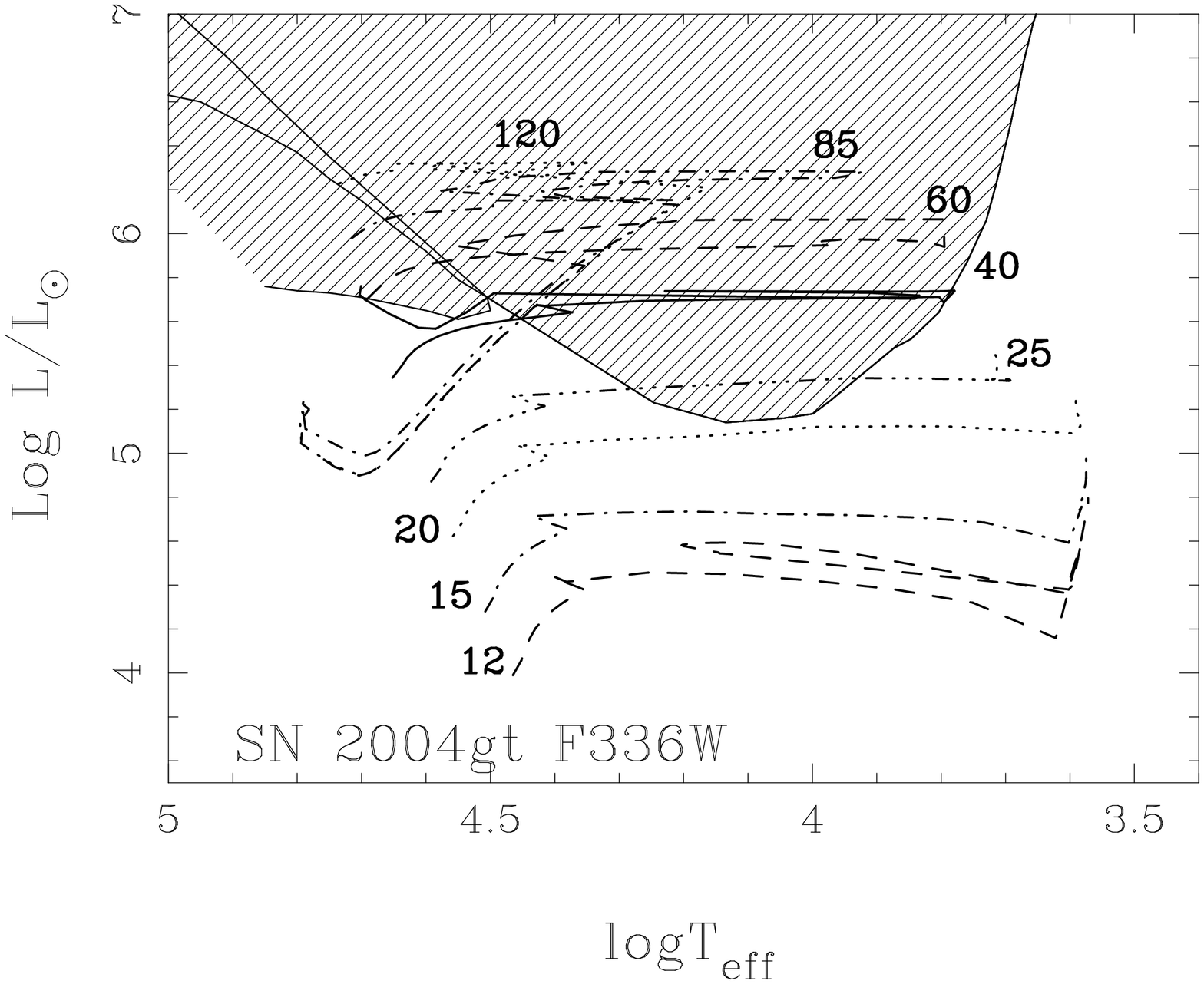}}
\rotatebox{-90}{\includegraphics[width=5cm]{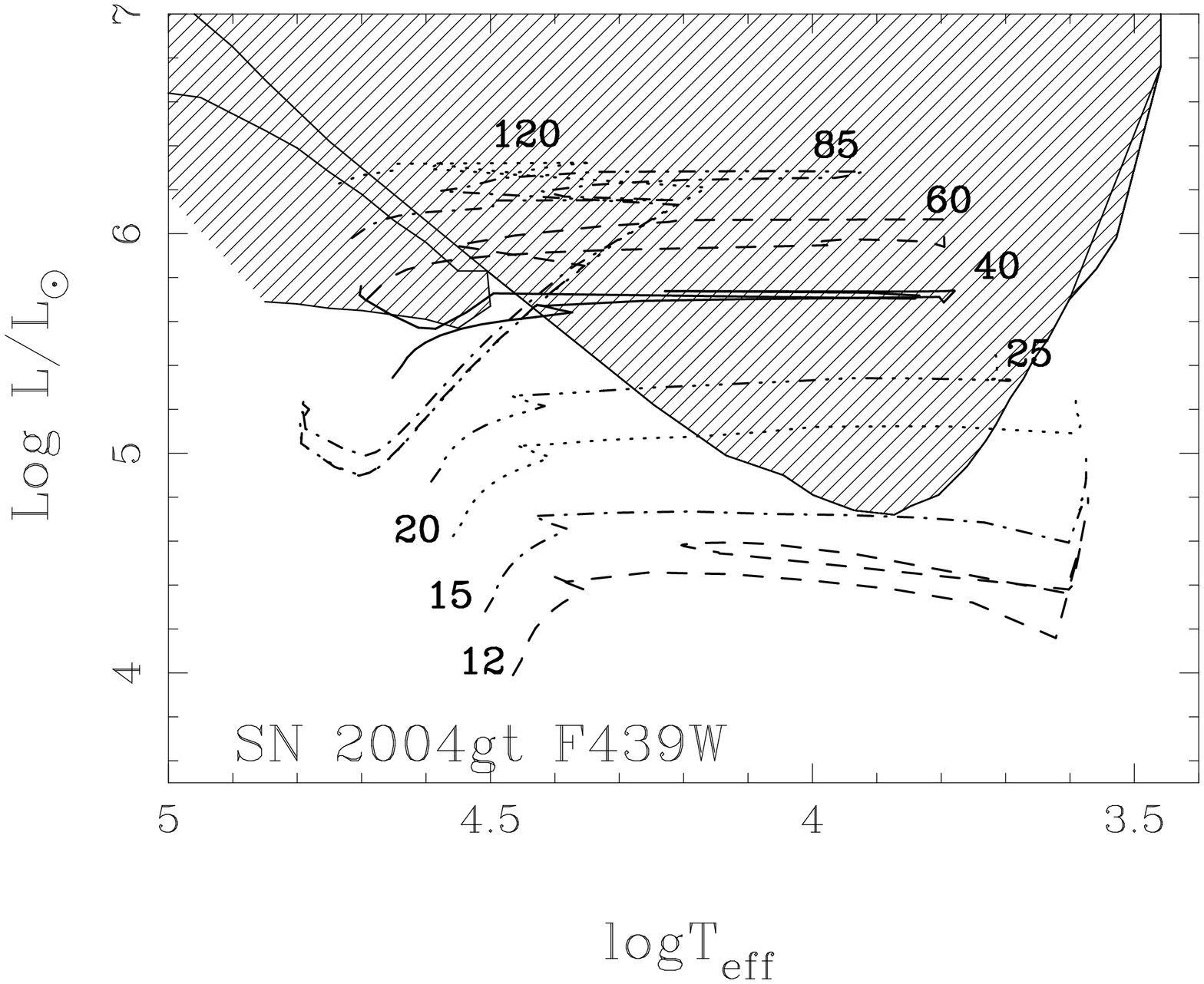}}
\rotatebox{-90}{\includegraphics[width=5cm]{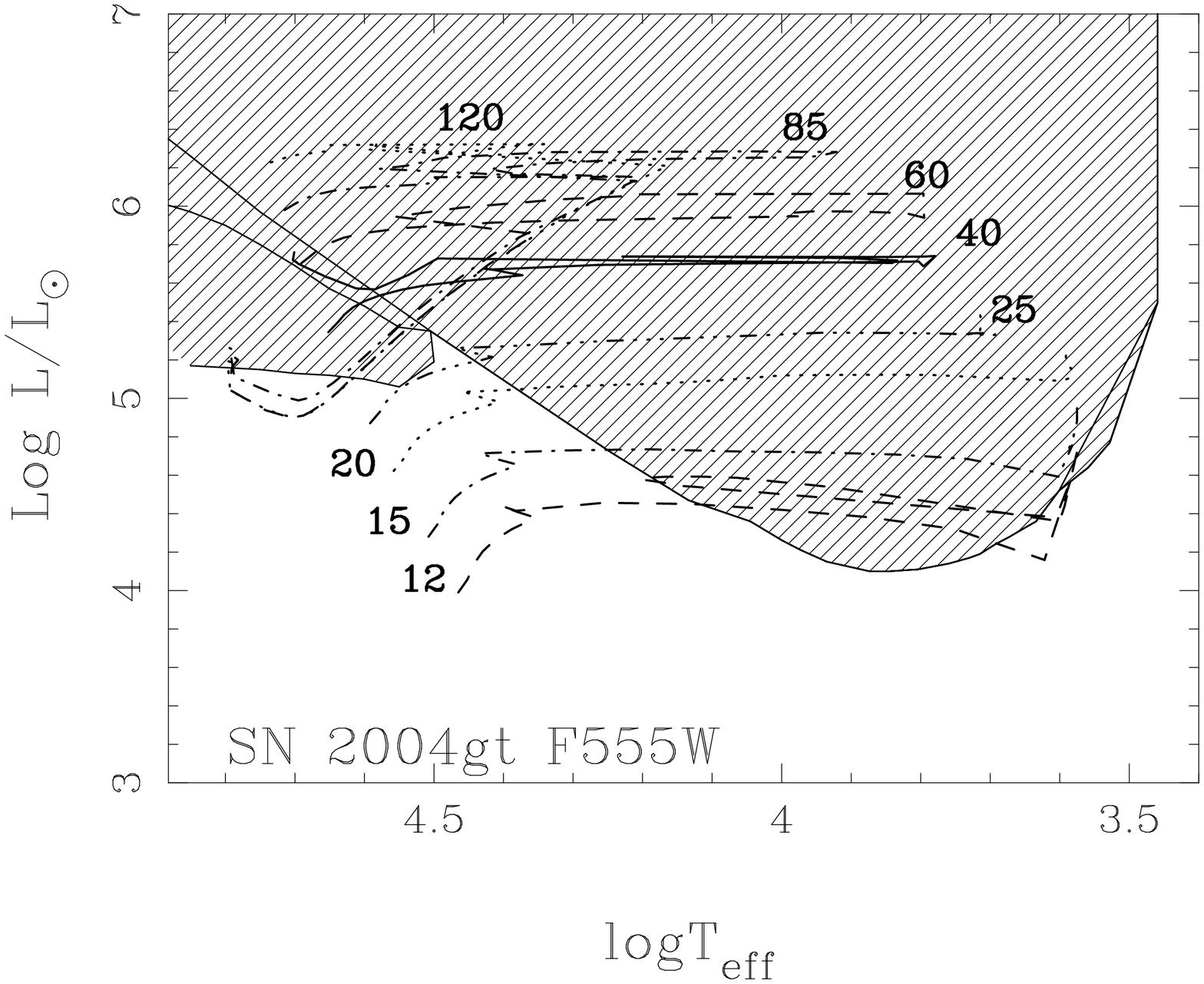}}
\rotatebox{-90}{\includegraphics[width=5cm]{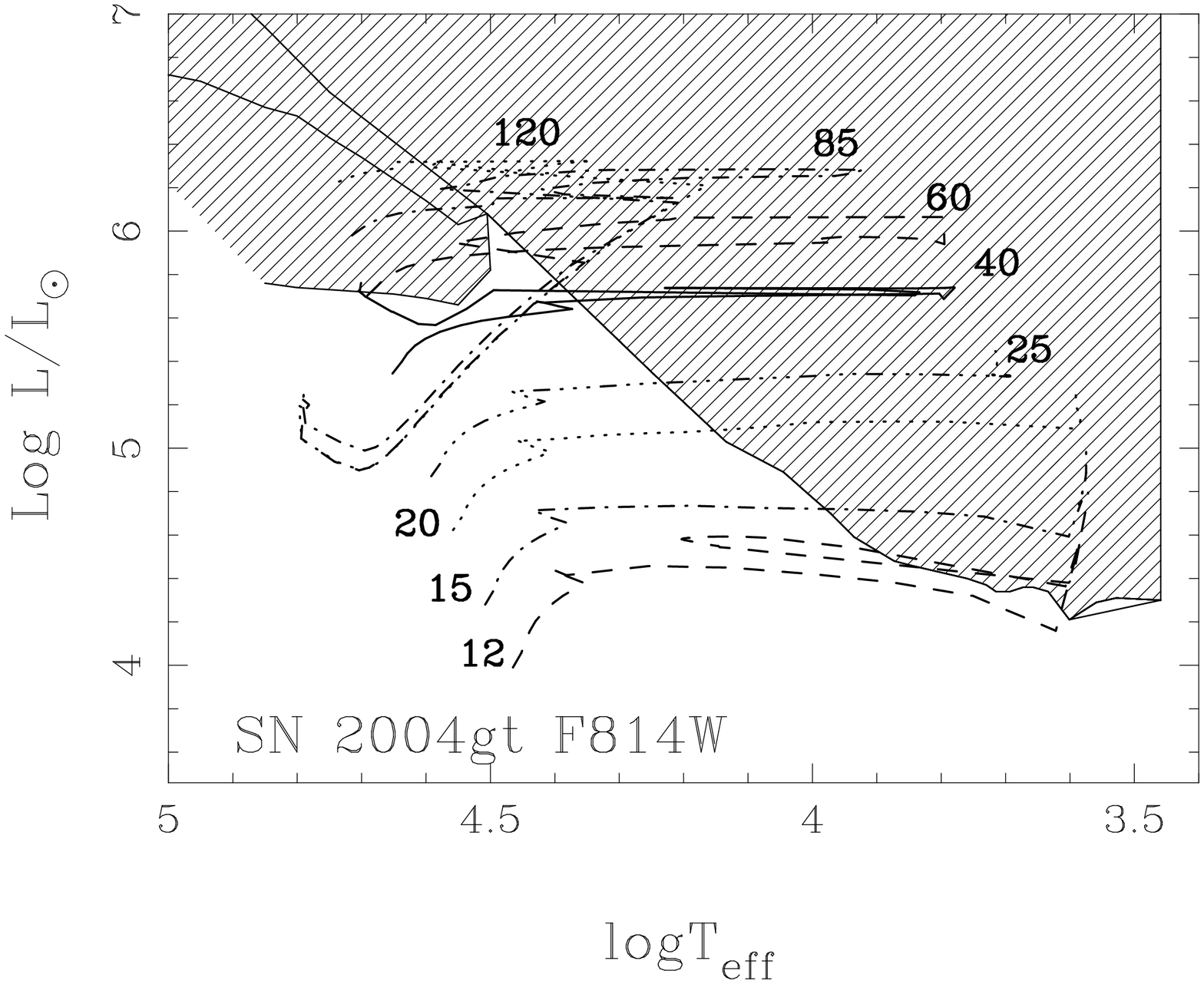}}
\caption{Hertzsprung-Russell diagrams showing solar metallicity stellar evolution tracks.  Overlaid are the detection thresholds for the pre-explosion HST WFPC2 F336W, F439W, F555W and F814W images.  A progenitor lying in the grey regions would have been significantly detected.}
\label{hrd}
\end{figure}

\citet{2005astro.ph..1323M}, \citet{smartt02ap} and \citet{vandykprog} have shown how the pre-explosion photometry may, in the event of the non-detection of the progenitor, be used to place mass limits on the progenitor.  The pre-explosion detection thresholds were converted to luminosity thresholds for a range of supergiant spectral types, using the colours, temperatures and bolometric corrections given by \citet{drill00}.  In addition synthetic photometry was conducted on model WR spectra \citep{wolfspec,2005astro.ph..1323M}, using the STSDAS packages {\it synphot}, to calculate the colours and bolometric corrections for these types of stars.  These colours and corrections are dependent on two parameters: the effective temperature and the ``Transformed Radius.''  \citet{2005astro.ph..1323M} discuss how the two-parameter space of model spectra was sampled to completely measure the spread in colours and bolometric corrections.  The principal consequence of this two-parameter dependence is that at high temperatures, when the progenitors are expected to be WR stars, the uncertainty in the luminosity increases with temperature compared to the relationship for supergiants.  The observed absolute magnitude detection limits were converted to luminosities using Eqtns. 6 and 7 of \citet{2005astro.ph..1323M}.  The detection thresholds for the four different filters were plotted on HR diagrams, shown in Fig. \ref{hrd}, and compared with stellar evolution tracks to estimate mass limits.\\   
The F336W  pre-explosion observation does not place a stringent constraint of the progenitor, missing both the red supergiant and blue WR end points of the stellar evolution tracks for low-intermediate and high-mass stars, respectively.  The F439W observation excludes yellow and red supergiant progenitors, for $M_{ZAMS}\approx20-25M_{\odot}$ as well as the lowest mass stars which end their lives with a WR phase ($M_{ZAMS}\approx25-40M_{\odot}$). The constraint against a WR progenitor, with initial mass of $40M_{\odot}$, is weak ,however, as the luminosity uncertainty is large.  The red supergiant progenitors for stars with $M_{ZAMS} <25M_{\odot}$ are completely excluded  by the F814W observation (including the location of the progenitor of the type IIP SN  2003gd; \citealt{smartt03gd}).  The F555W observation places much stronger constraints on both sides of the HR diagram.  The red supergiant progenitors for stars with $M_{ZAMS} <25M_{\odot}$ are completely excluded (including the location of the progenitor of the type IIP SN  2003gd; \citealt{smartt03gd}).  A much tighter constraint on the WR branch excludes a $M_{ZAMS}\approx25-40M_{\odot}$ progenitor.  All four pre-explosion  observations clearly indicate that the SN 2004gt was not the explosion of a super-massive $M_{ZAMS}\approx 120M_{\odot}$ star.  The depth of the observations, despite not being sufficient to detect the progenitor, provide useful limits on the progenitor of SN 2004gt.  The use of the smaller distance of $13.8\pm1.7$Mpc to NGC 4038, determined 
by \citet{2004AJ....127..660S}, would result in the threshold limits on 
the HR diagrams being lowered by $\mathrm{\sim0.28 log L_{\odot}}$.  The 
consequence of this would be to increase the likelihood of the detection of a WR star progenitor arising from stars with initial masses $\sim40M_{\odot}$.  This shift is, however, smaller than the uncertainties of the thresholds at the blue side of the HR diagram.  The pre-explosion observations do not exclude possible blue lower mass progenitors in binaries, with $M_{ZAMS}=20-40M_{\odot}$ as given by the age determined for the surrounding stellar population.  The evolution of such a progenitor would have been affected by angular-momentum and mass exchange with a companion, which would strip the progenitor's H envelope leaving a low-luminosity high-temperature He and C-O star as the progenitor \citep{podsiIbc04}.  \citet{2005astro.ph..1323M} have suggested this as a possible explanation for the lack of the detection of a number of progenitors of type Ib/c SNe.  This progenitor scenario was also invoked by \citet{nom94i} for the type Ic SN 1994I.  \citet{podsiIbc04} argue that prompt collapse of a single star would produce a faint SN, whereas SN 2004gt is of normal brightness.  The distance to NGC 4038, 19.2Mpc, is prohibitive to conducting a spectroscopic search for the companion \citep{maund93j}.  In Table\,\ref{Ibcstats} we compile all the directly measured upper limits for the magnitudes of the progenitors of SNe Ibc, in various pass-bands.  We can determine the upper final mass limit, i.e. the mass {\em prior} to explosion, of a WR star progenitor from the mass-luminosity relationship of the Langer models\citep{lang89}.  These upper limits are somewhat higher, although not inconsistent with the calculated final mass of the star which produced SN~2002ap ($5-7$\msol ; which includes $\sim$2\msol\ allowance for a compact remnant) determined from modelling the spectral evolution of the supernova \citep{mazz02}.  \citet{1990ApJS...73..685V} give the range of observed absolute magnitudes for Galactic WN stars as $M_{V}=-3.2$ to $8.0$.  Assuming a uniform distribution in absolute brightness implies that the pre-explosion observations are sensitive to $\approx 50\%$ of WN stars, mostly WNL stars.   The luminosities of WC stars in the LMC \citep{2002A&A...392..653C} are $\mathrm{log(L/L_{\odot})<5.8}$, which places them within the lower uncertainty boundary for the detection of WR stars (see Fig. \ref{hrd}). This shows that these observations cannot constrain the progenitor scenario by themselves.  A comparison of these final upper mass limits may, however, be used with SN evolution models and their mass budget to constrain a WR progenitor scenario.   Understanding the progenitors of Type Ic SNe, whether high-mass WR stars or low-mass stars in binaries, is important for understanding the relationship between SNe and GRBs \citep{hjo03dh}.  The high rate of SNe events in the merging galaxies NGC 4038/4039 ($\mathrm{0.2-0.3 \; SNe\; yr^{-1}}$; \citealt{2000AJ....120..670N}) in conjunction with the large amount of deep imaging of this system available suggests that NGC 4038/4039 is a good candidate for the detection of progenitors of Type Ibc SNe in the future.
\par
The location of SN 2004gt, in the galaxy NGC 4038, in pre-explosion HST WFPC2 images has been identified using differential astrometry with a post-explosion high-resolution HST ACS observation of SN 2004gt.  The progenitor location was identified with an accuracy of 5mas.  The progenitor was not detected in any of the deep F336W, F439W, F555W and F814W pre-explosion imaging.  The detection thresholds of the pre-explosion observations suggest that SN 2004gt arose from a single star progenitor with $M_{ZAMS}\gtrsim40M_{\odot}$ or a $M_{ZAMS}\approx20-40M_{\odot}$ star in a binary.  This is the most restrictive mass limit placed on any type Ib/c SN yet, and this is attributable to the depth of the available pre-explosion imaging.
\acknowledgements
Based on observations made with the NASA/ESA Hubble Space Telescope and obtained from the Data Archive at the Space Telescope Science Institute, which is operated by the Association of Universities for Research in Astronomy, Inc., under NASA contract NAS 5-26555. These observations are associated with programs GO-5962 and GO-10187.  JRM and SJS acknowledge financial support from PPARC.  FS acknowledges partial support from the NSF through grant AST02-05994.

\clearpage

\bibliographystyle{apj}
%\bibliography{/home/jrm/bibliography/main}

%%%%%TABLE1
\clearpage

\begin{table}
\caption{\label{Ibcstats} 
Compilation of all the directly determined limits to the progenitors of SNe Ibc.}
\begin{tabular}{llcccccrr}
\hline\hline
Supernova & Type  &    \multicolumn{5}{c}{Limiting magnitudes} &  $\log L_{\rm {bol}}$ & $M_{\rm Final}$\\
          &       & $M_U$ & $M_B$& $M_V$ & $M_R$ & $M_I$ & ($L_{\odot}$)  &  ($M_{\odot}$) \\
\hline
2000ds    &  Ib   &  ...  & ...  & -6.0  & ...   & -6.5  &   5.2 &  10  \\
2000ew    &  Ic   &  ...  & ...  & -6.4  & ...   & ...   &   5.5 &  15   \\
2001B     &  Ib   &  ...  & ...  & -8.0  & ...   & ...   &   6.0 &  25  \\
2002ap    &  Ic   & -8.7  & -6.8 & -7.2  & -7.8  & -8.4  &   5.6 &  18  \\
2004gt    &  Ic   & -8.42 & -6.82& -5.31 & ...   & -6.77 &   5.1 &  9   \\
\hline         
\end{tabular}
\\
NOTES.--- Limits are from this paper, \citet{smartt02ap}, and \citet{2005astro.ph..1323M}, with the corresponding bolometric luminosity limit calculated from \citet{2005astro.ph..1323M}.  The limiting magnitudes for SN~2002ap have been adjusted for the new large distance to M74 as determined by Hendry et al. (2005). The $M_V$ and $M_I$ magnitudes for SN~2000ds are HST $M_{F555W}$  and $M_{F814W}$; $M_V$ for SN~2000ew and SN~2001B are $M_{F606W}$ and $M_{F555W}$ respectively.  For SN 2004gt the limiting magnitudes $M_U$, $M_B$,  $M_V$ and $M_I$ are HST $M_{F336W}$,  $M_{F439W}$, $M_{F555W}$ and $M_{F814W}$ respectively.
\end{table}
%%%%%%%%%%%%%%%FIGURES%%%%%%%%%%%%%%%%%
\clearpage

\clearpage
\end{document}